\documentclass[final,5p,twocolumn,10pt]{elsarticle} 

\usepackage{lineno}
\usepackage{enumitem}
\modulolinenumbers[5]

\usepackage{amsmath}
\usepackage{amssymb}
\usepackage{stmaryrd}
\usepackage{amsthm}
\usepackage{amsfonts}
\usepackage{dsfont}
\usepackage{graphicx}
\usepackage{upgreek}
\usepackage{bm}
\usepackage{color}
\usepackage{float}
\usepackage{tikz-cd}

\usepackage{todonotes}

\usepackage{algorithm,algpseudocode} 
\usepackage{tikz,pgf}

\biboptions{numbers,sort&compress}

\DeclareFontFamily{OT1}{pzc}{}
\DeclareFontShape{OT1}{pzc}{m}{it}{<-> s * [1.200] pzcmi7t}{}
\DeclareMathAlphabet{\mathpzc}{OT1}{pzc}{m}{it}

\newcommand{\ba}{{\mathbf{a}}}
\newcommand{\bb}{{\mathbf{b}}}
\newcommand{\bx}{{\mathbf{x}}}

\newcommand{\bt}{{\mathbf{t}}}

\newcommand{\indic}{{\mathbf{1}}}

\newcommand{\dimm}{\mathrm{d}}

\newcommand{\obs}{\mathcal{O}}
\newcommand{\ont}{\mathcal{C}}
\newcommand{\fiber}{\mathpzc{f}}

\newcommand{\powerset}{\mathcal{P}}

\newcommand{\R}{{\mathds{R}}}

\newcommand{\SO}[1]{{\mathrm{SO}(#1)}}
\newcommand{\SE}[1]{{\mathrm{SE}(#1)}}

\newcommand{\conf}{{\mathsf{C}}}

\renewcommand{\th}{$^\text{th}$ }

\theoremstyle{definition}

\newcommand{\eq}[1]{(\ref{#1})} 
\newcommand{\com}[1]{} 

\usetikzlibrary{shapes} 
\usetikzlibrary{arrows,shapes,positioning}
\usetikzlibrary{decorations.markings}
\tikzstyle arrowstyle=[scale=1]
\tikzstyle directed=[postaction={decorate,decoration={markings,
		mark=at position .99 with {\arrow[arrowstyle]{stealth}}}}]
\definecolor{pinegreen}{cmyk}{0.92,0,0.59,0.25}
\definecolor{royalblue}{cmyk}{1,0.50,0,0}
\definecolor{lavander}{cmyk}{0,0.48,0,0}
\definecolor{violet}{cmyk}{0.79,0.88,0,0}
\tikzstyle{cblue}=[circle, draw, thin,fill=cyan!20, scale=0.4]
\tikzstyle{qgre}=[rectangle, draw, thin,fill=green!20, scale=0.8]
\tikzstyle{rpath}=[ultra thin, red, opacity=0.4] 
\tikzstyle{bpath}=[ultra thin, blue, opacity=0.05]
\tikzstyle{bbpath}=[ultra thin, blue, opacity=1]
\tikzstyle{bbbpath}=[ultra thick, blue, opacity=1]
\tikzstyle{legend_isps}=[text=black]

\journal{Symposium of Solid and Physical Modeling}

\begin{document}

\begin{frontmatter}

\title{Automatic Support Removal for Additive Manufacturing Post Processing}
 
\author{Saigopal Nelaturi, Morad Behandish, Amir M. Mirzendehdel, and Johan de Kleer}
\address{\rm Palo Alto Research Center (PARC), 3333 Coyote Hill Road, Palo Alto,
	California 94304  \vspace{-15pt}}

\begin{abstract}

An additive manufacturing (AM) process often produces a {\it near-net} shape
that closely conforms to the intended design to be manufactured. It sometimes
contains additional support structure (also called scaffolding), which has to be
removed in post-processing. We describe an approach to automatically generate
process plans for support removal using a multi-axis machining instrument. The
goal is to fracture the contact regions between each support component and the
part, and to do it in the most cost-effective order while avoiding collisions
with evolving near-net shape, including the remaining support components. A
recursive algorithm identifies a maximal collection of support components whose
connection regions to the part are accessible as well as the orientations at
which they can be removed at a given round. For every such region, the
accessible orientations appear as a `fiber' in the collision-free space of the
evolving near-net shape and the tool assembly. To order the removal of
accessible supports, the algorithm constructs a search graph whose edges are
weighted by the Riemannian distance between the fibers. The least expensive
process plan is obtained by solving a traveling salesman problem (TSP) over the
search graph. The sequence of configurations obtained by solving TSP is used as
the input to a motion planner that finds collision free paths to visit all
accessible features. The resulting part without the support structure can then
be finished using traditional machining to produce the intended design. The
effectiveness of the method is demonstrated through benchmark examples in 3D.

\end{abstract}


\end{frontmatter}


\section{Introduction}

Many AM processes require generating support structures to sustain the
manufactured part so that it does not collapse under its own weight (in polymer
AM), burn due to intensive laser power, or warp due to thermal residual stresses
(in metal AM) as the material is deposited throughout the process. For a given
design, support structures directly add to AM cost and engineers aim to minimize
required supports typically by optimizing the build direction. However,
selecting the build direction is usually influenced or even dictated by other
parameters such as material properties, feature size, accuracy, and so on. Thus,
at least for the foreseeable future, support structures are inevitable for
fabricating many industrial parts across numerous AM technologies.

\subsection{Subtractive Post-Processing of Near-Net Shapes}

Producing quality parts using AM often requires post-processing operations on
the {\it near-net} shape, including the AM part and the support structure. The
post-processing is typically in the form of machining or subtractive
manufacturing (SM) in general. Currently, post-process planning is performed
manually and is labor-intensive, especially in metal AM. Moreover, with the
advent of hybrid manufacturing technologies \cite{Yamazaki2016development}, the
entire workflow can be potentially automated for interleaved AM and SM
operations \cite{Behandish2018automated}, which requires accurate and efficient
process plans to generate as well as remove supports.

Engineers may be able to recommend a candidate set of support removal sequences
by visual inspection. For example, they quickly decide if a near-net shape has
support components that cannot be removed---a pathological example being when
supports are placed in internal voids. However, cutting tool assemblies are
complex constructions with indexable inserts, tool-holders, and optimized tool
shanks. Beyond simple cases, it is hard to manually plan the support removal
sequence, especially when there are many feasible solutions of marginally
different cost---which can make a difference for mass production.

The combination of the part, support, and tool assembly geometries define a
complex space of non-colliding tool \emph{configurations}, i.e., combinations of
translations and orientations. This free-space is a subset of the 6D
configuration space of rigid motions \cite{Lozano-Perez1983spatial}. At the
beginning of the post-process, only a subset of support components is accessible
through the free space, while others are occluded by the part and surrounding
support components. Every time a support component is removed, more components
become exposed and the free space constantly grows. Finding a feasible sequence
for removing all supports one after another is a challenging spatial planning
problem; because it requires navigating a {\it dynamic} free space against a
near-net shape that is updated at every step. This problem does not have an
automated solution as of today.

For the purposes of this paper, we are only interested in post-process planning
for a {\it given} support structure, i.e., one that is generated by any method
and given as input. Post-processing ranges from picking the supports by hand
using a chisel-like tool, to programming collision free tool-paths for a
CNC-mill to machine them off. In either case, the engineer often minimizes the
effort by fracturing only the contact region between each support component and
the additively manufactured part, so that at the end of the post-process the
part and supports can be disengaged from the base-plate of the AM machine. This
approach is efficient because it does not require machining the entire scaffold
to expose the part and therefore saves time while extending tool life. Our
strategy is to use this property to our advantage to navigate the gigantic
search space.

\begin{figure}
	\centering
	\includegraphics[width=0.4\textwidth]{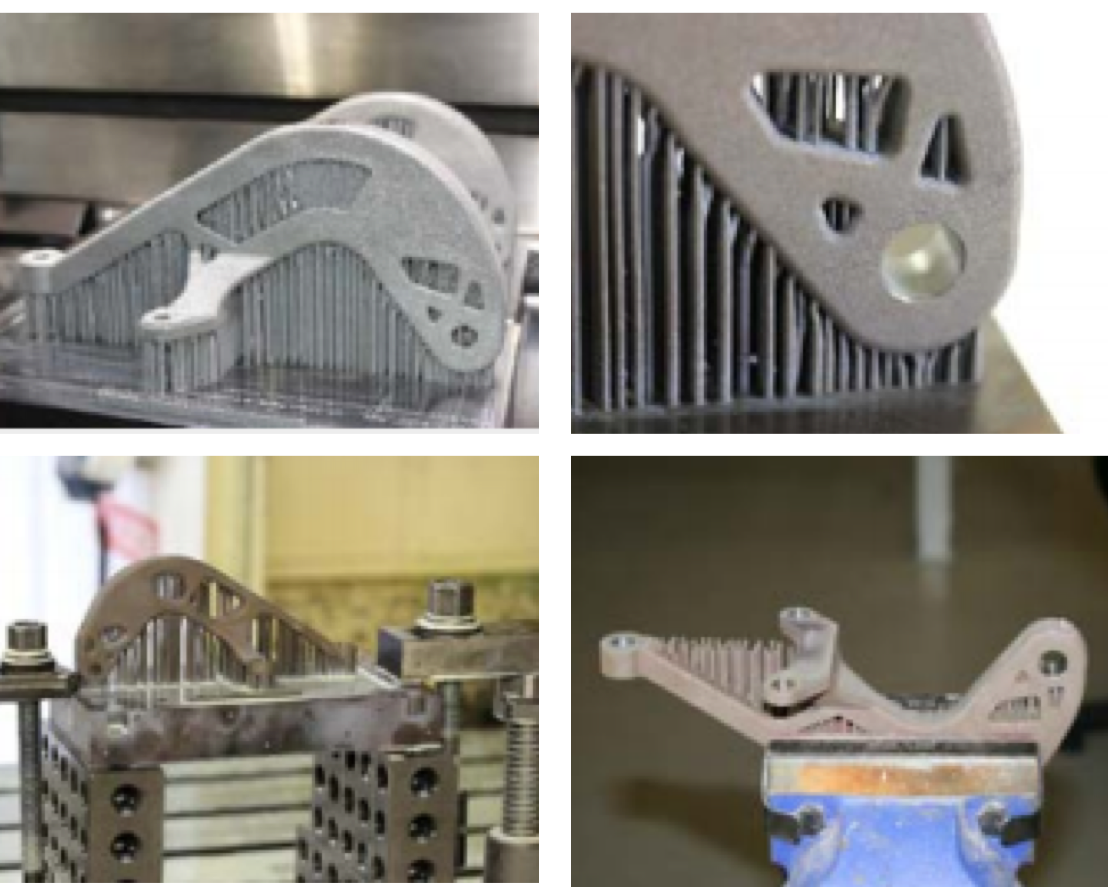}
	\caption{A part made using metal AM. The scaffolding is shown (top-left) along
	with a close-up (top-right). Two separate setups for SM are used for manual
	support removal (bottom-left and -right). The images are taken from
	\cite{Hamilton2016planning}.}
	\label{fig_metal}
\end{figure}

Figure \ref{fig_metal} illustrates a typical example of a metal AM part with
support structure.

\subsection{Related work}

The literature on support structures for AM is broad. Without attempting a
comprehensive review, we present some background on support structure reduction,
generation, and accessibility analysis. For a comprehensive review of the
influence of support materials on design, planning, and post-processing for AM,
see \cite{Jiang2018support}.

\subsubsection{Reducing Support Structures}

Support structures are sacrificial features that substantially add to
manufacturing cost by increasing build time, material usage, and
post-fabrication clean-up time. In metal AM, for instance, machine operation
costs (energy, hardware, and labor) are about 63\% of the total cost, material
costs are about 18\%, and post-processing costs are about 8\%, on average
\cite{Thomas2014costs}. Numerous researchers have investigated eliminating or
reducing support materials for AM from different perspectives. Support structure
layout is intimately coupled with the build direction. One of the most popular
manufacturing planning strategies in AM is to choose a build direction that
requires the least support material
\cite{Nezhad2010pareto,Das2015optimum,Paul2015optimization,Ezair2015orientation,Morgan2016part,Coupek2018reduction,Hehr2017five}.
It is also possible to design the part itself in such a way that it requires the
least amount of support while meeting other performance criteria. Multiple part
design optimization approaches focus on eliminating support structures
altogether by designing ``self-supporting'' shapes
\cite{Gaynor2016topology,Langelaar2016topology}. Others  optimize the design by
considering support minimization in a trade-off with competing objective
functions as a practical alternative
\cite{Mirzendehdel2016support,Qian2017undercut}.

\subsubsection{Generating Support Structures} 

Support structures must be designed to be accessible, heat-dissipating,
lightweight and load-bearing \cite{Jiang2018support}. Tree-like supports are a
popular design for metal AM that can be manually generated through commercial
packages such as \textsf{Meshmixer\textsuperscript{TM}}.  An automated approach
for creating such supports was proposed in \cite{Vanek2014clever}. For
powder-bed metal AM, a contact-free approach was developed in
\cite{Cooper2018contact} where the support structure is not attached to the
build part but effectively dissipates heat in the surrounding powder. For some
polymer AM processes, it is also possible to use water-soluble support materials
that can alleviate issues regarding support components that are hard-to-access
or connected to thin features \cite{Ni2017fabrication}. There are also recent
efforts towards generating dissolvable supports for metal AM that exploit
chemical and electrochemical stability between different metals
\cite{Hildreth2016dissolvable}. However most of the metal AM processes are based
on powder-bed fusion that are predominantly single-material.

\subsubsection{Accessibility Analysis}   

Since it is not always possible to completely eliminate the support structures
(or make them dissolvable, etc.), it must be ensured that all support components
are accessible according to a specified post-process. Little attention has been
paid to automating the accessibility analysis to remove supports, presumably
because the interacting geometries are usually complex. While the ability to
fabricate such complex shapes is one of AM's great strengths, it makes
post-process planning more challenging than, for instance, SM planning for
traditional feature-based CAD models. When the tool assembly can translate and
rotate at the same time, it tremendously adds to the complexity. Before
addressing these problems, we briefly overview some of the challenges in
accessibility analysis.

Accessibility analysis can be conveniently formulated in terms of the absence of
interference at key locations on the intermediate part. The shape and degrees of
freedom (DOF) of the moving tool substantially contribute to the difficulty of
accessibility analysis, which is why many approaches try to simplify one or both
of these inputs. For example, Spitz et al. \cite{Spitz1998accessibility} analyze
the accessibility of coordinate measuring machines (CMM) under translation in
terms of the free space (computed via a Minkowski sum) of part and CMM
instrument. CMMs often have elongated stems to probe the part and therefore
local accessibility analysis can be simplified in terms of visibility. In
general, visibility analysis is used commonly as an approximation of
accessibility analysis when tools can be abstracted into one or more rays
(ignoring the thickness and shape) \cite{Limaiem2000integrated,
	Arbelaez2008cleanability,Nelaturi2015automatic}. These approaches can be quite
effective for a variety of problems, but generalize poorly when the interference
is substantially influenced by tool geometry, as is the case with realistic tool
assemblies.

Analysis of contact and interfering configurations is a fundamental problem in
robot motion planning. Classical approaches focus on analytically describing the
configuration space ($\conf$-space) obstacle as the solution to polynomial
equations that capture all possible contact interactions (e.g., vertex-face,
face-face, etc.) between a polyhedral part/obstacle and tool/robot
\cite{Canny1988complexity}. When the interacting shapes are complex non-convex
polyhedra, exact computation of the $\conf-$space obstacle may require
intersecting thousands of primitive patches. For general spatial motions,
$\conf-$space mapping using analytical representations is typically restricted
to relatively simple geometries. For a comprehensive review of $\conf-$space
mapping, see \cite{Wise2000survey}.

Formulating motion planning problems in the configuration space of rigid motions
is a classical approach \cite{Lozano-Perez1983spatial} and has been studied
extensively. The key insight is to transform the dynamic problem of planning the
motion of a moving robot to the static problem of finding a path for a point in
the free space. while the formalism is elegant, computing 6D free spaces has
been considered a barrier to its practical implementation. Modern motion
planners rely on sampling-based approaches such as probabilistic roadmaps
\cite{Kavraki1994probabilistic,Kavraki1996probabilistic,Hsu98finding,Boor1999gaussian}
to locally sample collision-free configurations and concatenate collision-free
paths. This approach is quite successful, but requires explicitly specifying
start and goal states for the motion planner. Identifying these start and goal
states is itself a challenging spatial reasoning problem in support removal and
other manufacturing planning applications.

To summarize, most approaches to accessibility analysis are based on simplifying
either the interacting geometries, or the DOF available to the moving shape. In
this paper, we make no such restrictions. We use the formalism of $\conf-$space
modeling to formulate accessibility analysis for support removal, while using
efficient computations of the free space by taking advantage of the known
locations for finite and (presumably small) dislocation features. We show that
the approach scales to practical situations involving complex part and tool
geometries.

\subsection{Contributions \& Outline}

We introduce an automatic spatial planning approach to identify a feasible and
cost-effective sequence of support removal operations and a path to execute
them.

Our approach relies on an {\it explicit} computation of the sampled (dynamically
evolving) free space. It applies to parts, support structures, and tool
assemblies of arbitrary shape, and multi-axis machines that have simultaneous
access to translating and rotating DOF.

The paper is organized as follows:

In Section \ref{sec_form}, we formulate the accessibility analysis
problem in the $\conf-$space and outlines a recursive algorithm to identify a
sequence in which the supports can be removed from a near-net shape. A key idea
to make the computations on an evolving $\conf-$space tractable is to compute a
maximal collection of removable supports at every ``round'' of the recursive
algorithm, i.e., the collection of all supports whose every connection to the
part is accessible in at least one orientation (Section \ref{sec_removal}).
Brief introductions to $\conf-$space formulation and computations that are
required to identify the maximal collection are provided in Sections
\ref{sec_obs}  and \ref{sec_contact}. In particular, we model the set of
accessible orientations at the connections of each support component to the part
can be viewed as a `fiber' in configuration space obtained from a lifting
operation.

In Section \ref{sec_plan} we present algorithms to plan for the sequence in
which the supports of a given maximal collection are removed to optimize a cost
function. We formulate this task as a traveling sales problem (TSP)
\cite{Lin1965computer} on a graph whose nodes are represented by the fibers and
edges are weighted by the cost of traveling between them. Once a candidate plan
is generated, we invoke a standard motion planning algorithm to compute
collision-free tool-paths between fracture points to remove support components
for the current round. The recursive algorithm is repeated until no support
components remain.

Section \ref{sec_results} demonstrates nontrivial support removal on a
nontrivial 3D bracket and 3D tool that has access to all six DOFs. Results are
provided to underscore the practical implementation of the proposed algorithms.
\section{Identifying Removable Supports} \label{sec_form} 

Removing supports by cutting off their contact regions with the part is an
efficient alternative to machining the entire support material using a
traditional milling process. For the latter, in the same way that one would
remove material from a full raw stock
\cite{Nelaturi2015automatic,Behandish2018turning}, one can clean out the support
material by sweeping the tool within the support material over the {\it
	continuum} of accessible configurations. In our approach, we need only {\it
	finite} contact configurations to peel off the support components. This
assumption is critical in our ability to efficiently compute the free space for
an evolving near-net shape.

At the beginning, it is likely that only a subset of support components are
removable in this manner with a given tool. Intuitively, we may imagine a forest
of support components (e.g., columns) in a near-net shape, where the occluded
columns are inaccessible until the columns along the periphery are removed.

Our approach uses known properties of the $\conf-$space of relative rigid
transformations (detailed in Sections \ref{sec_obs} and \ref{sec_contact}) to
automatically identify the maximal collection of supports that are guaranteed to
be removable from a near-net shape in an intermediate state. Briefly, the
algorithm proceeds as follows: The $\conf-$space obstacle is computed explicitly
and all contacting but non-colliding configurations of the tool are extracted. A
support component is deemed removable iff all of its dislocation features are
accessible through {\it contact}, meaning that there exists at least one
configuration in the {\it boundary} of the $\conf-$space obstacle that
corresponds to each dislocation feature. All removable support components are
collected and designated for removal at the current ``round'' of the
post-process plan. After removing this collection from the near-net shape, the
$\conf-$obstacle reduces in size, more dislocation features become accessible,
hence more support components become removable. The algorithm recursively
identifies a sequence of removable supports that are peeled off by the cutting
tool to finally converge to the desired part.

Note that this process is independent of the precise order and tool-path%
\footnote{For the moment, we take the existence of such a path for granted,
	which is not always the case. We return to this issue in Section
	\ref{sec_ompl}.}
in which the collected support components of a given round are removed in
practice. A manufacturability test is also encoded in the algorithm, which can
identify early on if the near-net shape includes support components that are
inherently unreachable even after removing the outer layers.

\subsection{Constructing Configuration Space Obstacles} \label{sec_obs}

We formulate the support removal problem for $\dimm-$dimensional shapes
(typically, $\dimm=2$ or $3$). Throughout the paper, we will use illustrations
in 2D for building intuition, and show 3D results in Section \ref{sec_results}.

Queries about the interference of a moving body (e.g., the tool assembly) $T
\subseteq \R^\dimm$ against a stationary set of 3D obstacles (e.g., the near-net
shape) $N \subseteq \R^\dimm$ are conceptualized as evaluating membership
predicted against the $\conf-$space obstacle $\obs(N,T) \subseteq \conf$ which
partitions the $\conf-$space of relative rigid transformations $\conf :=
\SE{\dimm}$ into three disjoint regions representing free, contacting, and
interfering configurations respectively. These regions correspond to the
exterior, boundary, and interior of the $\conf-$space obstacle, respectively.

Constructing the obstacle $\obs(N,T)$ is a critical step in our algorithm.
Practical approaches use the fact that the particular group structure of $\SE{\dimm}
= \SO{\dimm} \rtimes \R^\dimm$ (i.e., the Lie group of rigid transformations)
allows it to be factored \cite{Selig2005geometrical} into two simpler and
lower-dimensional subgroups; namely, $\SO{\dimm}$ (for rotations) and $\R^\dimm$
(for translations). Hence, we can represent each rigid transformation $\tau \in
\SE{\dimm}$ as a tuple $\tau \equiv (r, \bt)$ with $\bt \in \mathbb{R}^\dimm$
and $r \in \SO{\dimm}$. In 2D, the former is a 2D vector while the latter is
represented by a angle $\theta \in [0, 2\pi)$. In 3D, the former is a 3D vector
while the latter can be represented in a number of different ways (e.g.,
orthogonal matrices, quaternions, axis-angle, Euler angles, etc.).

\begin{figure*}
	\label{fig_configs}
	\centering
	\includegraphics[width=0.9\textwidth]{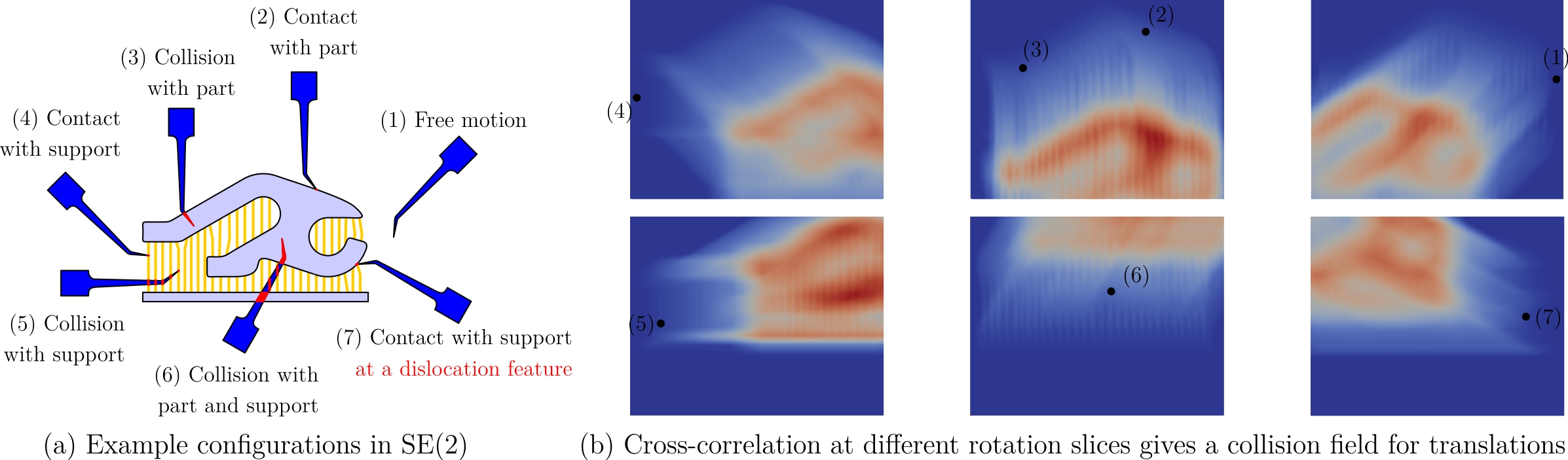}
	\caption{At every iteration of support removal planning, different tool
	configuration modes (e.g., free, contact, and collision) are determined by
	point membership classification (PMC) against the $\conf-$space obstacle. In
	this 2D example, at every tool orientation $r \in \mathrm{SO}(2)$ in (a), the
	overlap measure (i.e., area of the interference regions highlighted in red) for
	all possible translations is given by a convolution field over the $r-$slice in
	(b). For a given translation (annotated points on the $r-$slices), the value of
	the field determines the overlap measure.}
	\label{fig:illustration}
\end{figure*}

Let us denote a rotation of $T \subseteq \R^\dimm$ by $r \in \SO{\dimm}$ by $T_r
\subseteq \R^\dimm$. Assuming that we are only dealing with solids (i.e.,
compact-regular seminalaytic sets in $\R^\dimm$)
\cite{Requicha1980representations}, the $\conf-$obstacle $\obs(N,T)$ may be
expressed as follows:
\begin{equation}
	\obs(P,T) = \bigcup_{r \in \SO{\dimm}} \mathsf{i}(N \oplus (-T_r)),
	\label{eq_obs}
\end{equation}
where $\mathsf{i}(\cdot)$ represents the topological interior operator. The
pointset $(N \oplus (-T_r))$ is the topological closure of the translational
$\conf-$space obstacle of $N$ with $T_r$, i.e., the $\conf-$obstacle for a {\it
	fixed} orientation $r \in \SO{\dimm}$. It is well-known
\cite{Kavraki1995computation,Middleditch1988application,Latombe2012robot} that
this pointset may be calculated in terms of a Minkowski sum.%
\footnote{The Minkowski sum of two pointsets $A,B \subseteq \R^\dimm$ is defined
	as $(A\oplus B) = \{\ba + \bb ~|~ \ba \in A ~\text{and}~ \bb \in B \}$.}
The $\conf-$space for all rotations is thus obtained by unifying the different
translational $\conf-$spaces, each forming a slice of the $\conf-$obstacle. In
practice, $\obs(N,T)$ may be approximated as a {\it stack} of a finite number of
$r-$slices, indexed by rotations sampled in $\SO{\dimm}$. This approach is
commonly used to calculate $\conf-$space obstacles for planar motions
\cite{Latombe2012robot}.

Notice that $(N \oplus (-T_r))$ represents the closure of the translational
$\conf-$space obstacle and its boundary, denoted by $\partial(N \oplus (-T_r))$
represents all the translations of the moving body (fir the fixed orientation)
that cause surface contact but no volumetric interference with $P$. The contact
configurations $\ont(N,T) \subseteq \SE{\dimm}$ are therefore obtained as:
\begin{equation}
	\ont(P,T) = \bigcup_{r \in \SO{\dimm}} \partial(N \oplus (-T_r)).
\end{equation}
The contact space is normally a lower-dimensional submanifold of $\SE{\dimm}$,
which makes its computation challenging. In practice, we approximate $\ont(N,T)$
with a slightly ``thickened'' set, i.e., a narrow (but measurable) region
surrounding the boundary of the $\conf-$space obstacle.

\subsection{Querying Contact Configurations} \label{sec_contact}

Our approach uses the fact that the Minkowski sum $(A \oplus B) \subseteq
\R^\dimm$ of a pair of arbitrary solids $A, B \subseteq \R^\dimm$ corresponds to
the support (i.e., $0-$superlevel set) of the {\it convolution} $(\indic_A \ast
\indic_B)$ of the indicator functions $\indic_A, \indic_B: \R^\dimm \to \{0,
1\}$ of the two sets.%
\footnote{The indicator function of a pointset $\Omega \subseteq \R^\dimm$ at a
	query point $\bx \in \R^\dimm$ is defined as $\indic_\Omega(\bx) = 1$ if $\bx
	\in \Omega$ and $\indic_\Omega(\bx) = 0$ if $\bx \not\in \Omega$.}
The advantage of this approach is that the convolution may be implemented in
terms of Fourier transforms.%
\footnote{The Fourier transform of convolution is the same as the product of
	Fourier transforms. Hence, the convolution can be computed by two forward
	transforms, one pointwise multiplication in frequency domain, and an inverse
	transform.}
When the sets are sampled uniformly on a regular grid, the fast Fourier
transform (FFT) provides an efficient and scalable implementation of each
translational $\conf-$space obstacle (i.e, $r-$slice of $\obs(P,T)$):
\begin{equation}
	(r, \bt) \in \obs(N, T) \quad\text{iff}\quad (\indic_N \ast \indic_{-T_r})
	(\bt) > 0, \label{eq_obs_conv}
\end{equation}
A configuration $\tau \equiv (r, \bt)$ is inside the obstacle if the convolution
does not vanish for at least one orientation $r \in \SO{\dimm}$. Note that the
convolution of nonnegative functions is always nonnegative, hence the free space
is defined implicitly by and equality test:
\begin{equation}
	(r, \bt) \not\in \obs(P, T) \quad\text{iff}\quad (\indic_N \ast \indic_{-T_r})
	(\bt) = 0. \label{eq_free_conv}
\end{equation}
The convolution provides a field of {\it overlap measure} values between $P$ and
$T_r$ over the translational $\conf-$space. In other words, $(\indic_N \ast
\indic_{-T_r})(\bt) = \mu^\dimm[N \cap (T_r+\bt)]$ measures the volume of
intersection (if any) between $P$ and the moved body $T_\tau := (T_r + \bt) =
\{\bx + \bt ~|~ \bx \in T_r\}$. Here, $\mu^\dimm[\cdot]$ represents the Lebesgue
$\dimm-$measure (e.g., area for $\dimm = 2$ and volume for $\dimm = 3$)
\cite{lysenko2010group}. The queried configuration is in the $\conf-$space
obstacle (resp. free space) iff this measure is greater than (resp. equal to)
zero.

The contact space $\ont(P, T)$ consists of configurations at which the overlap
measure is {\it critically} zero, meaning that there exists an infinitesimal
change to the configuration that can make it nonzero. To approximate $\ont(N,
T)$ in a computable fashion, we implicitly define an `$\epsilon-$contact' space
using the following predicate:
\begin{equation}
	(r, \bt) \in \ont_\epsilon(N, T) \quad\text{iff}\quad 0 < (\indic_N \ast
	\indic_{-T_r}) (\bt) < \epsilon. \label{eq_ont_conv}
\end{equation}
Using a small the threshold $\epsilon > 0$, one may assume that
$\ont_\epsilon(N, T)$ closely approximates $\ont(N, T)$ from a measure-theoretic
standpoint.

To summarize, if the $r-$slices are computed and stored beforehand (as a finite
number of convolutions), it is possible to query the overlap measure in constant
time at any sampled translation and rotation by interpolation against the stack
of convolution fields. In particular, $\ont(N,T)$ is numerically approximated as
the set of all transformations $(r,\bt) \in \SE{\dimm}$ where $0 < (\indic_N \ast
\indic_{-T_r})(\bt) < \epsilon$ for some tolerable interference volume $\epsilon
> 0$.

Figure \ref{fig:illustration} illustrates in 2D how the computed stack of
convolutions for different $r-$slices for sampled rotations is used to classify
a given tool configuration as free, contact, or colliding. In 3D, the same exact
method applies, except that each convolution is a 3D field and the rotations are
samples in 3D using any number of standard uniform or pseudo-uniform/random
sampling methods \cite{Yershova2010generating}.

Algorithm \ref{alg_contact} describes a simple (and parallel) procedure to
compute the $\epsilon-$contact space for a finite sample $\{r_s\}_{1 \leq s \leq
	n_1} \subset \SO{\dimm}$ of orientations. The convolution is faster to compute
on a regular sample $\{\bt_s\}_{1 \leq s \leq n_2} \subset \R^\dimm$ of
translations using uniform FFTs \cite{Kavraki1995computation}, which requires a
regular axis-aligned sample (i.e., voxelization) of the part and rotated tool.
The procedure takes $O(n_1 n_2 \log n_2)$ which is very close to linear time in
the number of sampled configurations $n := n_1 n_2$. Note that this procedure is
called only once per round of the recursive outer-loop (Algorithm
\ref{alg_removal}). See the results in Fig. \ref{fig_obs} of Section
\ref{sec_results}.

\begin{algorithm} [ht!]
	\caption{Computing $\epsilon-$contact configurations.}
	\begin{algorithmic}
		\Procedure{Contact}{$N, T; n_1, n_2, \epsilon$}   
		\State $\ont_\epsilon(N,T) \gets \emptyset$
		\State $\Theta \gets \Call{SampleRot}{n_1}$ \Comment{e.g., Hopf fibration
		\cite{Yershova2010generating}.}
		\ForAll{$r \in \Theta$}
		\State $T_r \gets \Call{Rotate}{T, r}$ \Comment{Rotates and resamples.}
		\State $\mu_{r} \gets \Call{Conv}{N,{T_r}}$ \Comment{At once via FFTs
		\cite{Kavraki1995computation}.}
		\ForAll{$\bt \in \mathrm{Grid}(n_2)$}
		\State $\ont_\epsilon(N,T) \gets \ont_\epsilon(N,T) \cup \{ (\bt,r) \ |\ 0 <
		\mu_{r}(\bt) < \epsilon  \}$
		\EndFor
		\EndFor
		\State\Return{$\ont_\epsilon(N,T)$}
		\EndProcedure 
	\end{algorithmic} \label{alg_contact}
\end{algorithm}

Note that storing the entire 6D set of $\conf-$space obstacle $\obs(N,T)$ for 3D
parts is not practical. However, the contact space $\ont(N,T)$, approximated by
$\ont_\epsilon(N,T)$, is a sparse subset of the $\conf-$space that can be
precomputed and queried later in constant-time.

\subsection{Identifying Removable Support Components} \label{sec_removal}

In this Section, we reason about the tool assembly's free space to automatically
construct a maximal collection of removable support components. This collection
is defined as the set of all support components that may be removed by the tool
assembly without interfering with the near-net shape, except by the tool tip at
the fracture points.

Let $P, S \subseteq \R^\dimm$ denote the part and support structure,
respectively, which only contact over their common boundary denoted by $F := (P
\cap  S) = (\partial P \cap  \partial S)$. The initial near-net shape is
obtained as $N := (P \cup S)$.

\begin{figure} [ht!]
	\centering
	\includegraphics[width=0.4\textwidth]{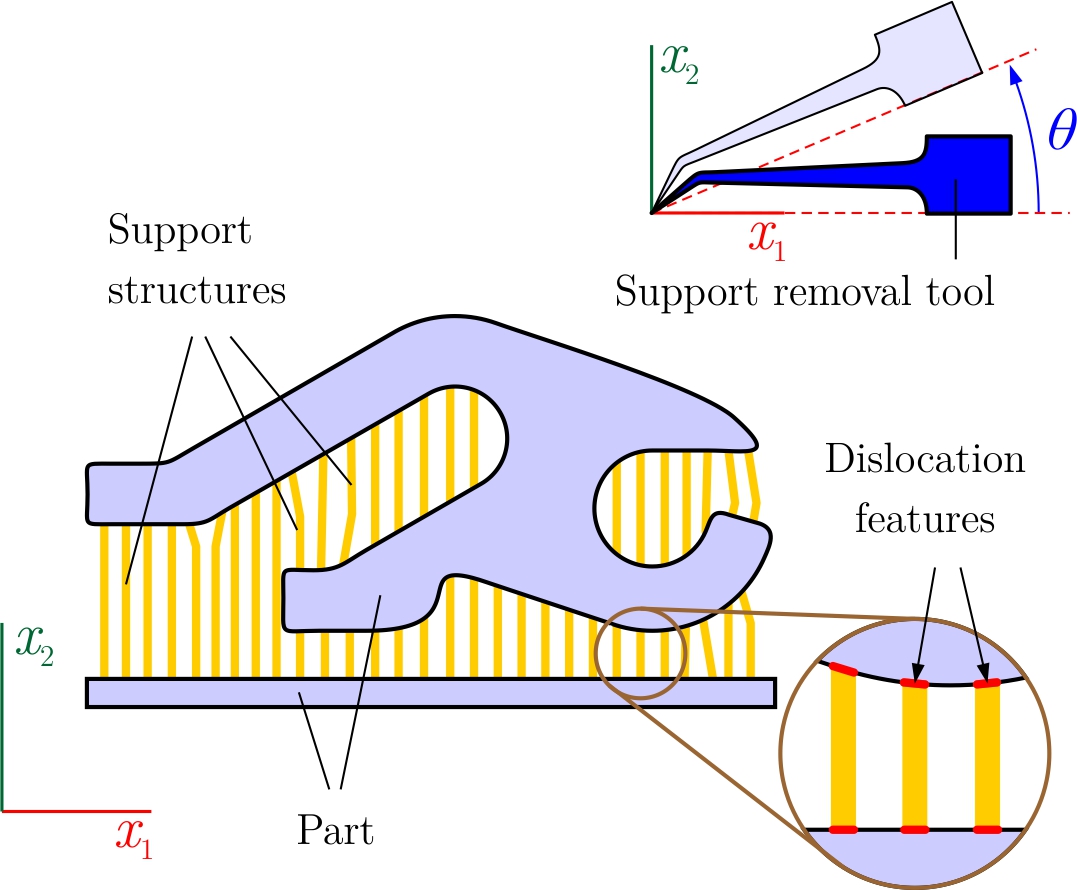}
	\caption{A 2D near-net shape, including the part and 
		support structure (i.e., scaffolding), and a support removal tool with three
	DOFs (two translations and one rotation). The $\conf-$space of rigid motions
	is $\SE{2}$ whose elements are parameterized by $(x_1,x_2,\theta)$.}
	\label{fig_nomenc}
\end{figure}

The support structure commonly consists of many connected components (called
hereafter {\it support components}):
\begin{equation}
	S = \bigcup_{1 \leq i \leq n_S} S_i, \quad(S_{i} \cap S_{i'}) = \emptyset
	\quad\text{if}~ i \neq i', \label{eq_components}
\end{equation}
where $S_i \subseteq \R^\dimm$ stands for the $i$\th component ($1 \leq i \leq
n_S$). We assume that a given support component is removable with a given
cutting tool if the tip of the tool can access and fracture all contact regions
between the support component and the part boundary in one or more
collision-free configurations, i.e., positions and orientations of the tool at
which the tool does not interfere with the part or other support components. The
contact region $F = (P \cap  S) = (\partial P \cap  \partial S)$ is also
decomposed into its connected components (hereafter called {\it dislocation
	features}):
\begin{equation}
	F = \bigcup_{1 \leq j \leq n_F} F_j, \quad(F_{j} \cap F_{j'}) = \emptyset
	\quad\text{if}~ j \neq j'. \label{eq_features}
\end{equation}
where $F_j \subseteq \R^\dimm$ stands for the $j$\th feature ($1 \leq i \leq
n_F$). Note that the indexing scheme for \eq{eq_components} and \eq{eq_features}
are different because every support component is connected to the part via two
or more dislocation features hence $n_F \geq 2 n_S$. The nomenclature is
illustrated in Fig. \ref{fig_nomenc} for a 2D example.

Support structures are typically designed to have small contact regions with the
part's surface to enable easier fracturing and removal and to minimize surface
roughness after removal. Note that sometimes support components are constructed
from one location on the part to another, but it is preferred that they are
constructed connect the part's surface to the machine's base plate. Moreover,
one support component can be connected to the part at multiple locations, e.g.,
when they are designed to branch out from a main trunk. To accommodate the most
general condition, we do not make any simplifying assumption on the shape,
connectivity, number, or layout of support components and their dislocation
features.

\begin{figure*}[ht!]
	\centering
	\includegraphics[width=0.9\textwidth]{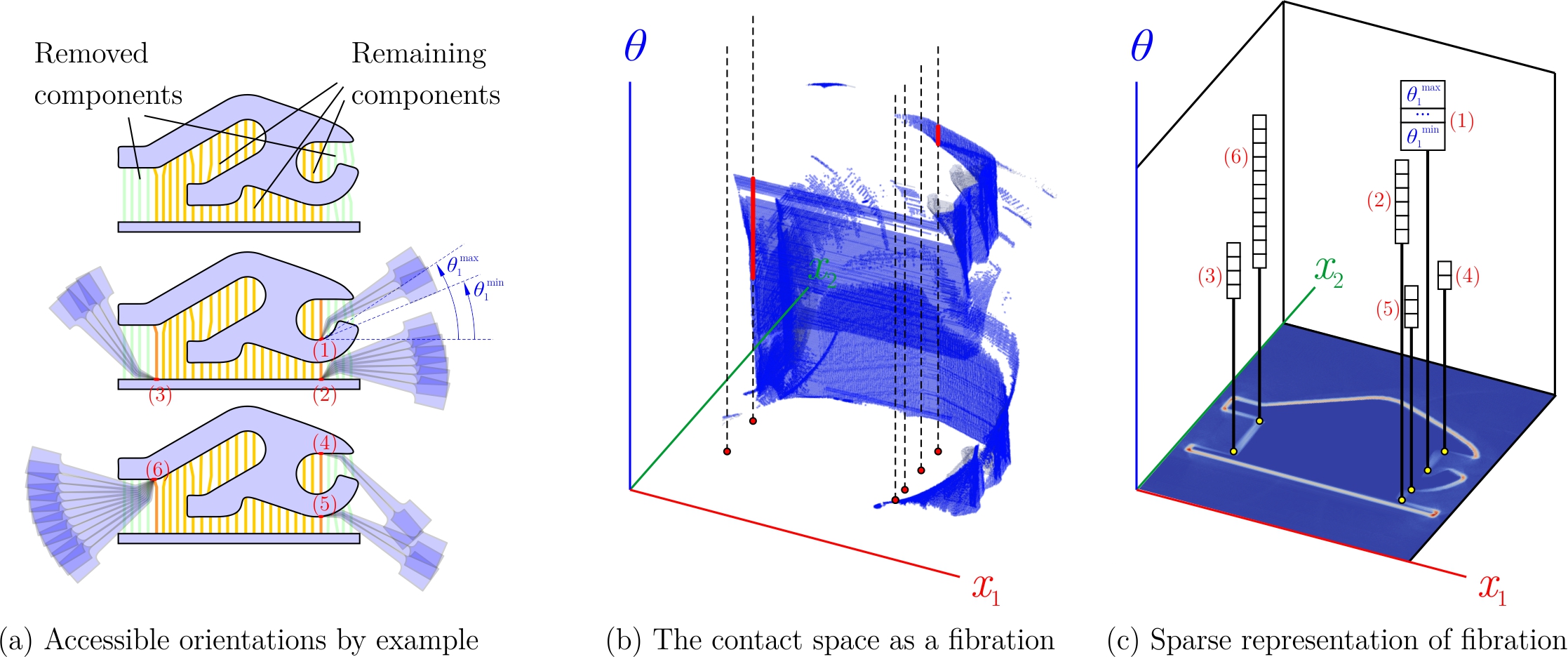}
	\caption{At every recursion of support removal planning, the accessible
	dislocation features (exemplified in (a)) are characterized by their inclusion
	in the the projection of the contact configurations. This implies the existence
	of a nonempty subset of orientations at which the feature's lifting by to the
	$\conf-$space (i.e., a `fiber') intersects the $\conf-$space obstacle's
	boundary. The resulting fibration is represented by a list of lists, each
	containing sampled orientations at which the dislocation feature is
	accessible.} \label{fig_fibers}
\end{figure*} 

Let us position the tool in its original configuration in such a way that the
cutter tip (or some point on the cutting surface) coincides with the origin of
the coordinate system used as a reference frame to quantify the rigid
transformations of the tool assembly $T \subseteq \R^\dimm$ with respect to the
near-net shape $N \subseteq \R^\dimm$ at a given round. This choice ensures that
a transformation $\tau \equiv (r, \bt) \in \SE{\dimm}$ brings the tool's cutter
(and not any other point on the tool) to the point $\bt \in \R^3$ in the
Euclidean $\dimm-$space where the near-net shape resides. Although this choice
is not necessary from a theoretical standpoint, it simplifies the accessibility
analysis by querying the pre-computed $\conf-$space map.

In a $\dimm-$dimensional space, the tool can operate with all $\dimm(\dimm+1)/2$
degrees of freedom, $\dimm(\dimm-1)/2$ of which are for rotations and the
remaining $\dimm$ are for translations. We can define \emph{projections} from
$\SE{\dimm} = \SO{\dimm} \rtimes \R^\dimm$ to $\SO{\dimm}$ and $\R^\dimm$,
respectively, that map every rigid transformation $\tau \equiv (r, \bt) \in
\SE{\dimm}$ to $r \in \SO{\dimm}$ and $\bt \in \R^\dimm$:
\begin{equation}
	\begin{tikzcd}[column sep=0.3em]
	& \SE{\dimm} \arrow[dl, "\pi_1"'] \arrow[dr, "\pi_2"] & \\
	\SO{\dimm} & & \R^\dimm
	\end{tikzcd}
	\begin{array}{l}
		\pi_1(\tau) = \pi_1(r, \bt) := r, \\
		\\
		\pi_2(\tau) = \pi_2(r, \bt) := \bt, 
	\end{array} \qquad\qquad\qquad
\end{equation}
Notice that the second projection depends on the earlier choice of origin.
Clearly, these maps are not invertible as functions. Nonetheless, we can define
{\it liftings} that take every $r \in \SO{\dimm}$ and $\bt \in \R^\dimm$ to a
{\it subspace} of $\SE{\dimm}$:%
\footnote{$\powerset(\Omega) = \{ \Omega' ~|~ \Omega' \subseteq \Omega \}$
	denotes the powerset of a set $\Omega$.}
\begin{equation}
	\begin{tikzcd}[column sep=0.1em]
		& \powerset(\SE{\dimm}) & \\
		\SO{\dimm} \arrow[ur, "\pi_1^{-1}"] & & \R^\dimm \arrow[ul, "\pi_2^{-1}"']
		\end{tikzcd}
		\begin{array}{l}
			\pi_1^{-1}(r) = \big\{ (r, \bt) ~|~ \bt \in \R^\dimm \big\}, \\
			\\
			\pi_2^{-1}(\bt) = \big\{ (r, \bt) ~|~ r \in \SO{\dimm} \big\}, \\
		\end{array}
\end{equation}
By extension, we can define set functions; for example, $\pi_2^{-1}:
\powerset(\R^\dimm) \to \powerset(\SE{\dimm})$ such that $\pi_2^{-1}(F)$ assigns
a copy of the entire $\SO{\dimm}$ to every point $\bt \in F$:
\begin{equation}
	\pi_2^{-1}(\Omega) := \{\pi_2^{-1}(\bt) ~|~ \bt \in F\} \cong (\SO{\dimm}
	\times F).
\end{equation}

Given a dislocation feature $F_j \subset P$, the set of all configurations at
which it can be touched by the tool tip can be obtained in terms of the above
projection/lifting maps:
\begin{equation}
	\fiber_j := \fiber(F_j, N, T) := \pi_2^{-1}(F_j) \cap \ont(N, T).
\end{equation}
In other words, every dislocation feature is lifted to the $\conf-$space by
pairing it with all possible orientations. The pairing results in the set of all
possible configurations that bring the tool tip to the dislocation feature (at
different orientations). Among them, the ones that are contained within the
contact space $\ont(N, T)$ are collision-free, hence they represent the set of
all accessible configurations. Hereafter, we refer to $\fiber_j \subset \SE{\dimm}$
as the {\it contact fiber} for the dislocation feature. Once again, for
computational purposes, we can approximate the contact fibers via
$\epsilon-$fibers, using the measurable $\epsilon-$contact region:
\begin{equation}
	\fiber_j \approx \fiber_\epsilon(F_j, N, T) := \pi_2^{-1}(F_j) \cap
	\ont_\epsilon(N, T).
\end{equation}
In practice, the dislocation features are small enough to assume that touching
{\it any} point on their circumference is sufficient to fracture it. As a
result, we can simply project the fiber from $\SE{\dimm}$ to $\SO{\dimm}$ to
collect all accessible orientations at all positions within the feature:
\begin{equation}
	\fiber_j^\ast := \fiber^\ast(F_j, N, T) = \pi_1 \big( \pi_2^{-1}(F_j) \cap
	\ont(N, T) \big).
\end{equation}
whose $\epsilon-$fiber approximations are given as expected by:
\begin{equation}
	\fiber_j^\ast \approx \fiber^\ast_\epsilon(F_j, N, T) = \pi_1 \big(
	\pi_2^{-1}(F_j) \cap \ont_\epsilon(N, T) \big).
\end{equation}
Figure \ref{fig_fibers} illustrates the fibers for a 2D example in which the
dislocation features are contracted to points. From an algorithmic perspective,
each $\epsilon-$fiber can be hashed into a list of sampled orientations anchored
at a different dislocation features. {\it A dislocation feature is accessible
	iff its fiber is nonempty}. As a result, we observe that a given support
component is removable iff every one of its dislocation features has a nonempty
fiber.

\begin{figure*}[ht!]
	\centering
	\includegraphics[width = 0.83\textwidth]{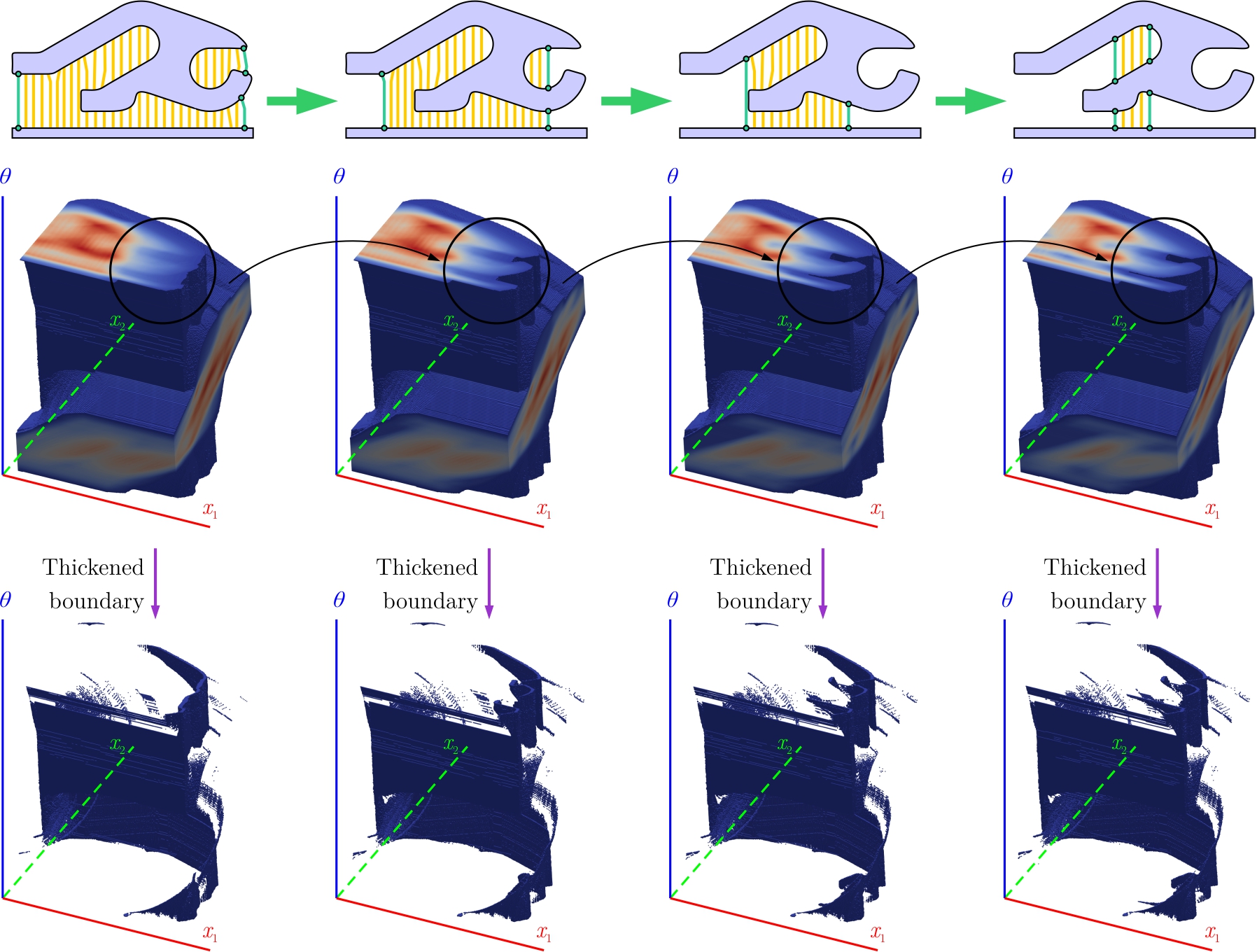}
	\caption{As the near-net shape goes through Algorithm \ref{alg_recursive}, the
	$\conf-$space obstacle keeps shrinking in size and more support structures
	become removable. At each recursion, the accessible dislocation features are
	identified by their fibers, exposed at the $\ont_\epsilon(N,T)$.}
	\label{fig_iterations}
\end{figure*}

Algorithm \ref{alg_removal} describes the procedure for finding the contact
fibers in a single round of recursive support removal planning. The procedure
performs $O(n_F)$ queries on every one of the convolution fields precomputed and
stored for $n_1$ sampled orientations, resulting in a total of $O(n_F n_1)$
queries---each convolution-read taking constant time. This analysis assumes that
we need $O(1)$ queries per each dislocation feature, because each feature is
sampled by one or at most a few point(s), which is reasonable. Every orientation
that is accessible is pushed into a stack that represents the fiber for the
queried feature.

\begin{algorithm} [ht!]
	\caption{Finding contact fibers.}
	\begin{algorithmic}
		\Procedure{Fibration}{$P, S, T, n_1, n_2, \epsilon$}
		\State $N \gets P \cup S$ \Comment{Current near-net shape.}
		\State $\{S_i\}_{1 \leq i \leq n_S} ~\gets \Call{CC}{S}$ \Comment{Support
		components.}
		\State $\{F_j\}_{1 \leq j \leq n_F} \gets \Call{CC}{P \cap S}$
		\Comment{Dislocation~ features.}
		\ForAll{$1 \leq j \leq n_F$}
				\State $\fiber^\ast_j \gets \emptyset$
				\Comment{~Initialize contact fibers.}
				\State $\ont_\epsilon(N, T) \gets \Call{Contact}{N, T, n_1, n_2, \epsilon}$
				\If{$\pi_2(\ont_\epsilon(N, T)) \cap F_j \neq \emptyset$}
					\State $\fiber^\ast_j \gets \fiber^\ast_j \cup \pi_1(\pi_2^{-1}(F_j) \cap
					\ont_\epsilon(N, T))$
				\EndIf
		\EndFor
		\State\Return{$\{\fiber^\ast_j\}_{1 \leq j \leq n_F}$}
		\EndProcedure 
	\end{algorithmic} \label{alg_removal}
\end{algorithm}
 
Note that before performing the queries, we can shortlist the $n_F$ features
rapidly to accessible ones by projecting the $\epsilon-$contact space to the Euclidean
$\dimm-$space (i.e., losing orientation information) and finding its common
points with the dislocation features:
\begin{equation}
	F_j ~\text{is accessible iff}~ \pi_2(\ont(N, T)) \cap F_j \neq \emptyset.
\end{equation}
This means that there exists at least one point in the dislocation feature $F_j$
that belongs to the projected contact space. The projection ensures that there
exists at least one non-colliding orientation at which the said point is
accessible. Note that the projected contact space can be obtained by unifying
the translational contact spaces $\partial(N \oplus (-T_r))$ for different
orientations:
\begin{equation}
\pi_2(\ont(N, T)) = \bigcup_{r \in \SO{\dimm}} \partial (N \oplus (-T_r)),
\end{equation}
Once again, the contact space can be approximated by the $\epsilon-$contact
criterion. As such, the above {\it early test} for accessibility can be rapidly
computed by summing up the indicator functions of the translational contact
spaces, which, in turn, are obtained as the $(0, \epsilon)-$interval level set
of the convolution fields for different $r-$slices. The resulting pointset is a
single field over the $\dimm-$space with a narrow support, containing all
positions in the vicinity of the near-net shape's boundary that can be accessed
by at least one orientation. For every point $\bt \in \R^\dimm$, we accumulate
the convolution value over $r-$slices for which the condition $0 < (\indic_N
\ast \indic_{-T_r})(\bt) < \epsilon$ holds. This incurs no additional
computation cost because as the convolutions are precomputed, their $(0,
\epsilon)-$interval level sets can be extracted and accumulated on-the-fly. The
result is a field with a narrow-band support over the $\dimm-$space that
quantifies non-colliding orientations of a given point. A 2D example of the
accumulated and projected field of contact measures is illustrated in Fig.
\ref{fig_fibers} (c) (base field). A dislocation feature is accessible iff it
has at least one point at which this function is nonzero. Note that this early
test counts the accessible orientations; however, it does not tell us {\it
	which} orientations (i.e., the fiber).

\begin{algorithm}[h!]
	\caption{Identifying removable supports.} \label{alg_recursive}
	\begin{algorithmic}
		\Procedure{Removable}{$P, S^\mathtt{t}, T, n_1, n_2, \epsilon$} 
			\If{$S = \emptyset$} \Comment{Base case: all supports removed.}
				\State\Return{$\emptyset$} \Comment{Final round: success.}
			\EndIf
			\State $\{\fiber^\ast_j\}_{1 \leq j \leq n_F} \gets \Call{Fibration}{P, S, T,
			n_1, n_2, \epsilon}$
			\State \Comment{Obtain nonempty contact fibers.}
			\State $\mathbf{I}^\mathtt{t} \gets \Call{NonEmpty}{\{\fiber^\ast_j\}_{1 \leq
			j \leq n_F}$} \Comment{Defined in \eq{eq_indices}.}
			\If{$\mathbf{I}^\mathtt{t} = \emptyset$}
				\State\Return{$\perp$} \Comment{Final round: failure.}
			\EndIf
			\State $S^\mathtt{t+1} = S^\mathtt{t} - \bigcup_{i \in \mathbf{I}^\mathtt{t}}
			S_i$ \Comment{Remaining supports.}
			\State\Return $\big\langle\mathbf{I}^\mathtt{t}, \Call{Removable}{P,
			S^\mathtt{t+1}, T, n_1, n_2, \epsilon} \big\rangle$
			\State \Comment{Appending sequence.}
		\EndProcedure
	\end{algorithmic}
\end{algorithm}

\subsection{Recursive Support Removal Rounds}

It is likely that multiple support components are removable from the near-net
shape at a given intermediate state. Intuitively, every round of the recursive
algorithm peels off all removable components to expose a new set of components.
The process is repeated until either all supports are removed, or the part is
deemed non-manufacturable because some supports are inaccessible. See the
results in Fig. \ref{fig_forest} of Section \ref{sec_results}.

At round$-\mathtt{t}$ of the support removal algorithm,%
\footnote{Note that superscripts here are used to indicate the rounds of the
	algorithm, and should not be confused with the subscripts used earlier to
	indicate connected components.}
the near-net shape is $N^{\mathtt{t}} := (P \cup S^{\mathtt{t}})$ with initial
conditions $N^{\mathtt{0}} = N$ and $S^{\mathtt{0}} = S$. Both $N^{\mathtt{t}}$
and $S^{\mathtt{t}}$ are monotonically reduced (in terms of set containment)
from one round to the next, i.e., $N^{\mathtt{t+1}} \subseteq N^{\mathtt{t}}$
and $S^{\mathtt{t+1}} \subseteq S^{\mathtt{t}}$.

Let $\mathbf{I}^\mathtt{t}$ represent the indices for the maximal collection of
removal supports at a given round, i.e., for every $i \in
\mathbf{I}^\mathtt{t}$, the support component $S_i \subseteq S^\mathtt{t}$ is
removable at round$-\mathtt{t}$. If $\mathbf{J}(i)$ represents the indices of
the dislocation features for the same support component, then $i \in
\mathbf{I}^\mathtt{t}$ iff for every $j \in \mathbf{J}(i)$, the fiber $\fiber_j$
is nonempty, i.e.,
\begin{equation}
	\mathbf{I}^\mathtt{t} = \big\{ 1 \leq i \leq n_S ~|~ \forall j \in
	\mathbf{J}(i): \fiber_j \neq \emptyset \big\}. \label{eq_indices}
\end{equation}
The remaining support for the next round is computed as: $S^{\mathtt{t+1}} =
S^{\mathtt{t}} - \bigcup_{i \in \mathbf{I}^\mathtt{t}} S_i$. The algorithm
continues until either of two termination criteria can happen:
\begin{enumerate}
	\item $N^{\mathtt{t}} = P$, i.e., $S^{\mathtt{t}} = \emptyset$, which implies
	successful removal of the entire support; or
	\item $N^{\mathtt{t}} = N^{\mathtt{t+1}}$ and $S^{\mathtt{t}} =
	S^{\mathtt{t+1}}$ which indicates that the remaining support components cannot
be reached.
\end{enumerate}
Figure \ref{fig_iterations} illustrates a few rounds of recursive support removal
for a 2D example. The $\conf-$space obstacle and contact space change as more
supports are removed.

\subsection{Necessary and Sufficient Conditions} \label{sec_mfg}

\begin{figure}[ht!]
	\centering
	\includegraphics[width=0.46\textwidth]{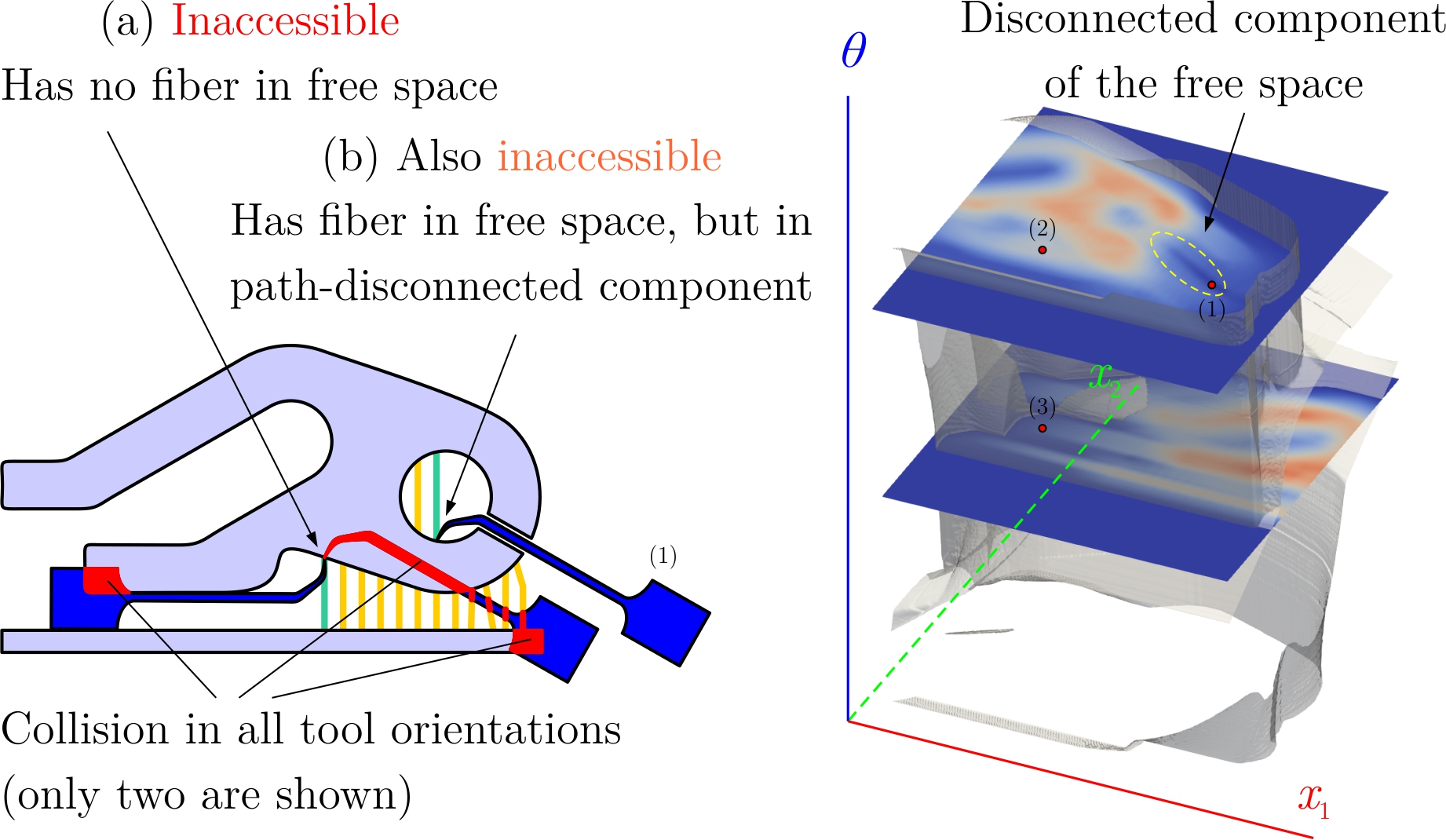}
	\caption{The test 
		provided by Algorithm \ref{alg_removal} is a necessary but not sufficient
		condition for manufacturability because it does not consider the
		path-connectedness of the free configuration space. If the free space is not
		path-connected, it is possible (but rare) for dislocation features to have a
		collision-free orientation that is not accessible through a collision-free
		path from an initial configuration.} \label{fig_inaccess}
\end{figure}

Algorithm \ref{alg_recursive} illustrates an implementation of the recursive
algorithm. All {\it potentially} accessible support components are identified by
checking if $\pi_2(\ont_\epsilon(N, T)) \cap F_j \neq \emptyset$ for every $j
\in \mathbf{I}_i$ for every remaining support component $S_i \subseteq
S^\mathtt{t}$. If at any given round, no supports are identified for removal,
the algorithm returns a failure message indicating that the support structure is
inherently not removable. In other words, this test provides a {\it necessary}
condition for the AM post-process to be feasible. While the test is quite useful
in finding the order in which the support components should be fractured, it
does not provide a {\it sufficient} condition. For example, the test does
account for the situations where there is a collision-free final configuration
to touch the dislocation feature, but there may not exist a collision-free {\it
	path} in the $\conf-$space that brings the tool from its initial configuration
to the final pose. This situation can happen if the  free space $(\obs(N,T))^c$
is not path-connected, and the initial and final configurations are located in
different connected components. Figure \ref{fig_inaccess} illustrates an
example of a false-positive. Such scenarios are rare if the support structure is
designed properly.

Although analyzing path-connectedness is important, $\SE{\dimm}$ is
high-dimensional for $\dimm \geq 3$ and the free space topology can be highly
complex. As opposed to reasoning about the topology, we can use a sampling-based
motion planner (e.g., based on probabilistic roadmaps
\cite{Kavraki1994probabilistic,Kavraki1996probabilistic,Hsu98finding,Boor1999gaussian}) to find a collision-free path from a reasonable starting configuration in $(\obs(N,T))^c$ to at least one configuration in the fiber.
\section{Support Removal Planning} \label{sec_plan}

In Section \ref{sec_removal} we have shown how an ordered sequence $\big\langle
\mathbf{I}^\mathtt{0}, \mathbf{I}^\mathtt{1}, \ldots, \mathbf{I}^\mathtt{t},
\ldots \big\rangle$ of the indices for maximal collections of removable support
components may be computed. Each $\mathbf{I}^\mathtt{t}$ represents a collection
of support components that can be removed {\it in any order} as long as all the
support components preceding them in the sequence have been removed in earlier
rounds of the algorithm.

In this section, we describe approaches to plan for the removal of the support
components within each group $\mathbf{I}^\mathtt{t}$ to reduce a cost function.
At the outset, we observe that any two collections $\mathbf{I}^{\mathtt{t}_1}$
and $\mathbf{I}^{\mathtt{t}_2}$ may be processed independently, because their
$\conf-$space obstacles are defined with respect to distinct near-net shapes
$N^{\mathtt{t}_1}, N^{\mathtt{t}_1}$. In other words, the order in which the
support components are removed at one round does not affect the order for
another round. As far as we assume an additive cost function, we may solve the
planning problem for different rounds in parallel.

\subsection{A Traveling Salesman Formulation}

Given a near-net shape $N^\mathtt{t} \subseteq \R^\dimm$, the tool assembly $T
\subseteq \R^\dimm$, and a sub-collection of support components consisting of
$\{S_i ~|~ i \in \mathbf{I}^\mathtt{t}\}$, removing all
$|\mathbf{I}^\mathtt{t}|$ components amounts to finding a collision-free path in
the $\conf-$space that moves the tool tip to all dislocation features that
connect the said support components to the part. These dislocation features are
given by $\{F_j ~|~ j \in \mathbf{J}_i ~\text{and}~ i \in \mathbf{I}^\mathtt{t}
\}$. The accessible configurations for every one of them is given by a fiber;
hence, the planning problem can be formulated as finding a path in the
$\conf-$space that visits all fibers one after another. It suffices for the path
to intersect with every fiber at least once, meaning that the tool touches the
dislocation feature in at least one accessible orientation.

The cost function for a candidate path can be defined in several plausible ways.
If we use machining time as the criteria and assume an almost constant speed in
the $\conf-$space, the path length in the $\conf-$space may serve as a cost
function. However, it is likely in practice to have different cost factors
associated with translational and rotational machine DOF. Here are two extreme
cases:
\begin{itemize}
	\item If we have a $3-$axis machine that needs re-fixturing  for every change
	of orientation, moving in the rotation space is significantly costlier than
	moving in the translational space. In this case, the cost function is to
	minimize the total number of orientations needed to visit all fibers. Using the
	mental picture in Fig. \ref{fig_fibers}, we may formulate this problem as
	finding the minimal number of $r-$slices to ``cut'' the given set of fibers,
	where each fiber is cut at least once.
	\item If for any reason changing orientations is cheaper, the problem is
	dramatically simplified. We can formulate the path planning conveniently in the
	Euclidean $\dimm-$space, especially since every fiber is projected to a small
	region (i.e., dislocation feature) which can be approximated by a point. The
	translational distance between the points, obtained by measuring the curve
	length in the $\dimm-$space serves as a cost function.
\end{itemize}

Consider a pair of configurations $\tau_1 \in \fiber_{j_1}$ and $\tau_2 \in
\fiber_{j_2}$, each selected from a different fiber that needs to be visited at
a given round, i.e., $j_1 \in \mathbf{J}_{i_1}$, $j_2 \in \mathbf{J}_{i_2}$,
$i_1, i_2 \in \mathbf{I}^\mathtt{t}$, and $j_1 \neq j_2$. The curve-length of a
geodesic in the $\conf-$space the connects these configurations is given by the
Riemannian distance $\| \ln(\tau_1^{-1} \tau_2) \|_2$ in which $\ln(\cdot)$
stands for the standard logarithm function,%
\footnote{The logarithm function on a Lie group defines a local coordinate
	system over a neighborhood by mapping the group elements to the Lie algebra
	(i.e., the tangent vector space) \cite{Selig2005geometrical}.}
which can be computed directly on the matrix representation of the relative
transformation $\tau_1^{-1} \tau_2 \in \SE{\dimm}$, and $\| \cdot \|_2$ is the
Frobenius (i.e., $L_2$) norm.  The minimum distance between all pairs of
configurations from the two fibers is a lower-bound to the length of any curve
segment that connects the two fibers in the $\conf-$space:
\begin{equation}
	\mathsf{d}(\fiber_{j_1}, \fiber_{j_2}) := \min_{\tau_1 \in
	\fiber_{j_1}}\min_{\tau_2 \in \fiber_{j_2}} \| \ln(\tau_1^{-1} \tau_2) \|_2.
	\label{eq_riemannian}
\end{equation}
The above distance function is well-defined for every pair of fibers (as a
whole). We can construct an undirected complete graph that has
$|\mathbf{I}^\mathtt{t}|+1$ vertices, one for each fiber and an additional one
for the reference configuration from which the tool assembly starts moving and
returns to after removing supports. We can associate the above distance function
with every edge, noting that it provides an admissible heuristic to approximate
the actual cost function {\it before} motion planning. This is important for
computational tractability of the algorithm.

Finding the shortest path to visit all fibers (i.e., vertices or ``cities'') at
least once is thus formulated as a traveling salesman problem (TSP)
\cite{Lin1965computer}. Our goal is to create a Hamiltonian cycle (a tour) with
minimum cost, obtained by adding the costs of individual edges on the graph,
starting and ending at the reference vertex.

The TSP is NP-complete, but we may use an approximation algorithm
\cite{Cormen2009introduction} if the cost-function is a proper metric. Distance
functions defined via minima of pairwise distances as in \eq{eq_riemannian} do
not normally qualify because they may not satisfy reflexivity and triangle
inequality, even though the pairwise distance itself is a metric, as in the case
of the Riemannian metric. The more appropriate formulation in this case would be
to associate every configuration (as opposed to every fiber) with a graph
vertex, use the pairwise Riemannian distance, which is a proper metric, as the
cost associated with every edge, and solve a more complex generalization of TSP
in which every {\it group} of cities (i.e., fibers) has to be visited at least
once.

Fortunately, if the dislocation features are small and ignore the cost of moving
along a single fiber---i.e., in-place rotations at a given dislocation feature
before moving to the next feature---we can still formulate and solve the problem
as a TSP. Using the approximation algorithm in \cite{Cormen2009introduction}, we
may assert that the cost of the tour is no more than twice that of the minimum
spanning tree's weight. The algorithm will find a sequence of graph vertices
(configurations) that are visited exactly once. The vertex sequence then
delineates the start and goal configurations for a sequence of motion planning
problems that can be solved by a standard motion planner. Note that the contact
fiber configurations will include minimal interference with (corresponding to
cutting) the supports. The start and end configurations are locally perturbed to
ensure a collision free path can be found.

\subsection{Motion Planning over the Fibration} \label{sec_ompl}

We use the Open Motion Planning Library (\textsf{OMPL}) \cite{Sucan2012open} to
compute collision-free paths to go from one fiber to the next in the sequence
prescribed by the TSP solution. The start and goal configurations on every fiber
are selected using a {\it greedy} policy as follows: For the first
configuration, we start from a given initial pose of the tool. For every fiber
that is visited next in the sequence, we select the configuration on the fiber
that is the nearest (in terms of the Riemannian metric) to the preceding
configuration selected on the previously visited fiber in the sequence.

As discussed earlier, there may be cases when the free space is not
path-connected. In such cases, even after successful determination of removable
support components with nonempty fibers (via Algorithm \ref{alg_recursive}) and
ordering them via TSP solution, the motion planner may not be able to find a
collision-free path to a dislocation feature.

We note that having $\obs(N,T)$ explicitly computed could help find
collision-free paths by overloading the collision checker of the motion planner
with a constant-time membership query against $\obs(N,T)$. However, as pointed
out in Section \ref{sec_obs}, it is impractical to store an explicit
representation of the entire $\obs(N,T)$. The stored contact space $\ont(N,T)$
approximated by $\ont_\epsilon(N,T)$ represents a boundary of the $\conf-$space
obstacle under fairly general conditions, hence can be used as a substitute.

However, membership queries against boundary representations might be more
time-consuming than highly-optimized collision detection algorithms. Moreover,
once the start and goal configurations are known, motion planning via
probabilistic roadmaps is hard to beat in terms of both efficiency and accuracy.
In particular, the accuracy of membership queries against precomputed
$\obs(N,T)$ (even when it is possible, e.g., in 2D) is limited by the
rasterization resolution. We can use a conservative policy, meaning that the
discretized 3D models are chosen to strictly contain the ``exact'' CAD models to
ensure that the approximation to $\obs(N,T)$ obtained via discrete convolutions
also contains the exact $\conf-$space obstacle. While this is safe policy as it
overestimates collisions, it commonly results in false-positives, especially at
contact configurations, e.g., detecting a collision when the tool is touching
the dislocation feature while there is none.

The convolution field is a fast and effective approach to computing the
collision measures for a large number of configurations {\it at once}---unlike
collision detection on B-reps, which is better suited for a {\it single}
configuration at-a-time. It is critical in our ability to sift through a large
number of dislocation features and decide their accessibility quickly at every
round of the recursion with evolving near-net shape. However, after the
accessible configurations, motion planning can be done with collision detection
between the master CAD models (e.g., B-reps) with a much higher precision than
that of rasterized models. 
\section{Results} \label{sec_results}

\begin{figure}
	\centering
	\includegraphics[width=0.46\textwidth]{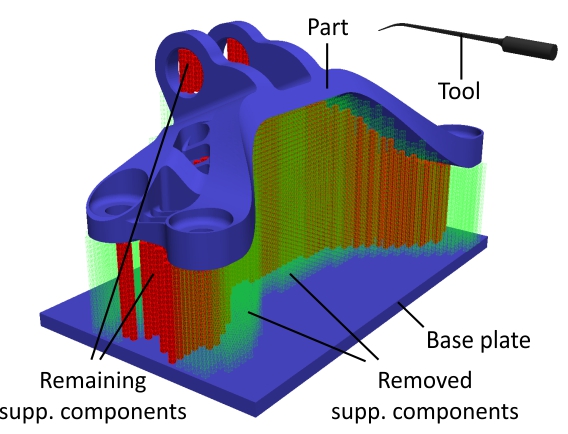}
	\caption{A near-net shape highlighting the AM part (blue), support removal tool
		(red), support structure (white), and dislocation features (green). The support
		components must be peeled off to fully disengage the part without leaving
		printed material behind.} \label{fig_setup}
\end{figure}

\begin{figure}[ht!]
	\centering
	\includegraphics[width=0.48\textwidth]{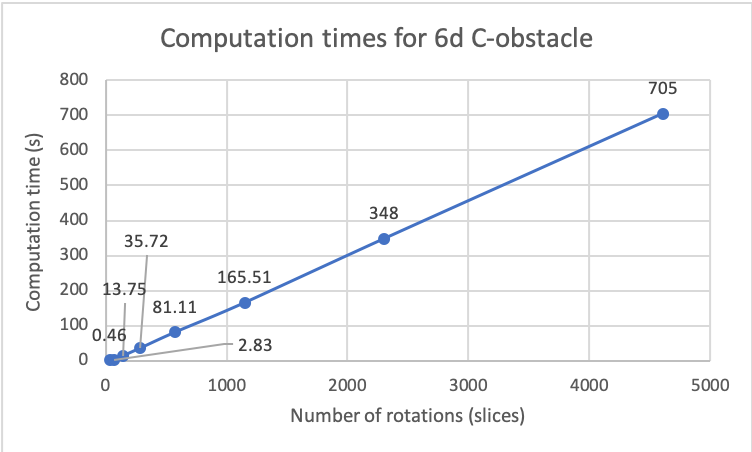}
	\caption{Computation times for the 6D$\conf-$space obstacle by sampling
		rotations and computing convolutions at a resolution of $512^3$ to represent
		each $r-$slice of the obstacle. The computations are performed using FFTs via
		\textsf{ArrayFire} on an \textsf{NVIDIA GTX 1080} GPU (2,560 CUDA cores, 8GB
		RAM) via \textsf{OpenCL}.} \label{fig_obs}
\end{figure}

\begin{figure}[ht!]
	\centering
	\includegraphics[width=0.48\textwidth]{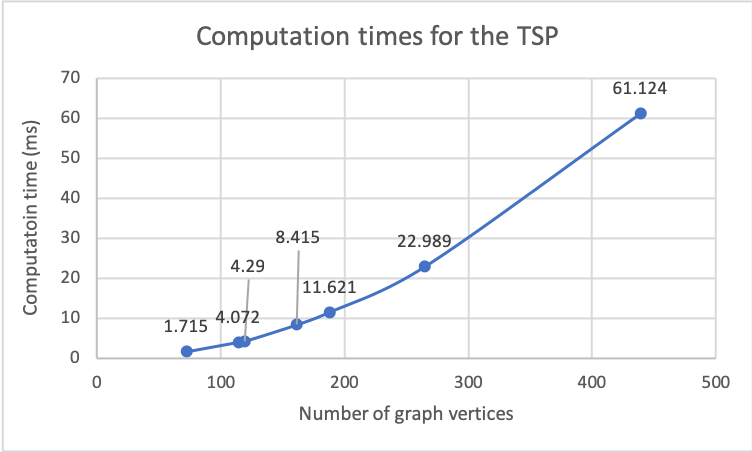}
	\caption{Computation times for the TSP in configuration space using the \textsf{Boost}
		Graph Library.} \label{fig_tsp}
\end{figure}

\begin{figure*}
	\centering
	\includegraphics[width=0.9\textwidth]{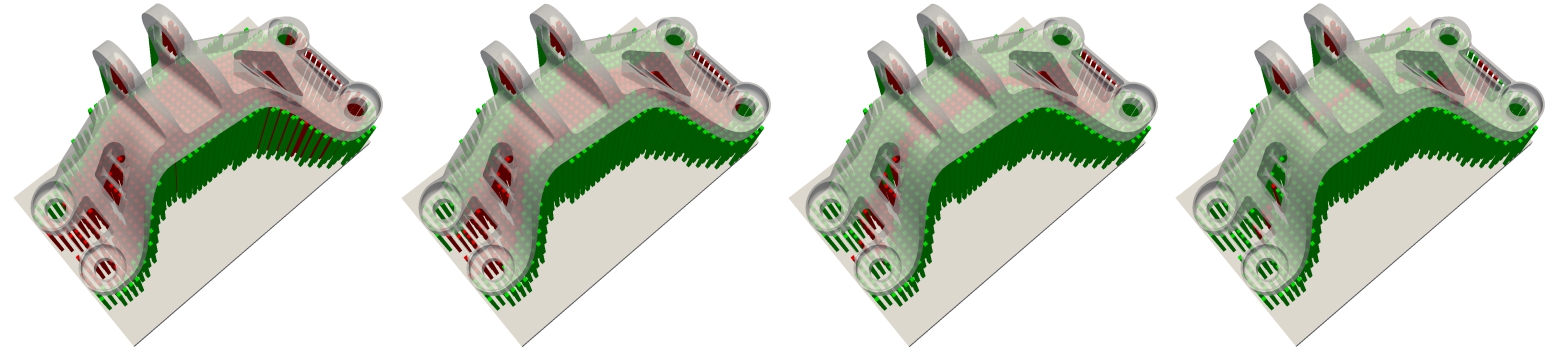}
	\caption{Several rounds of support removal via Algorithm
		\ref{alg_recursive}. At each round, a subset of the support components is
		connected to the part and the base plate by dislocation features that are
		accessible in at least one orientation.} \label{fig_forest}
\end{figure*}

\begin{figure*}[ht!]
	\centering
	\includegraphics[width=0.9\textwidth]{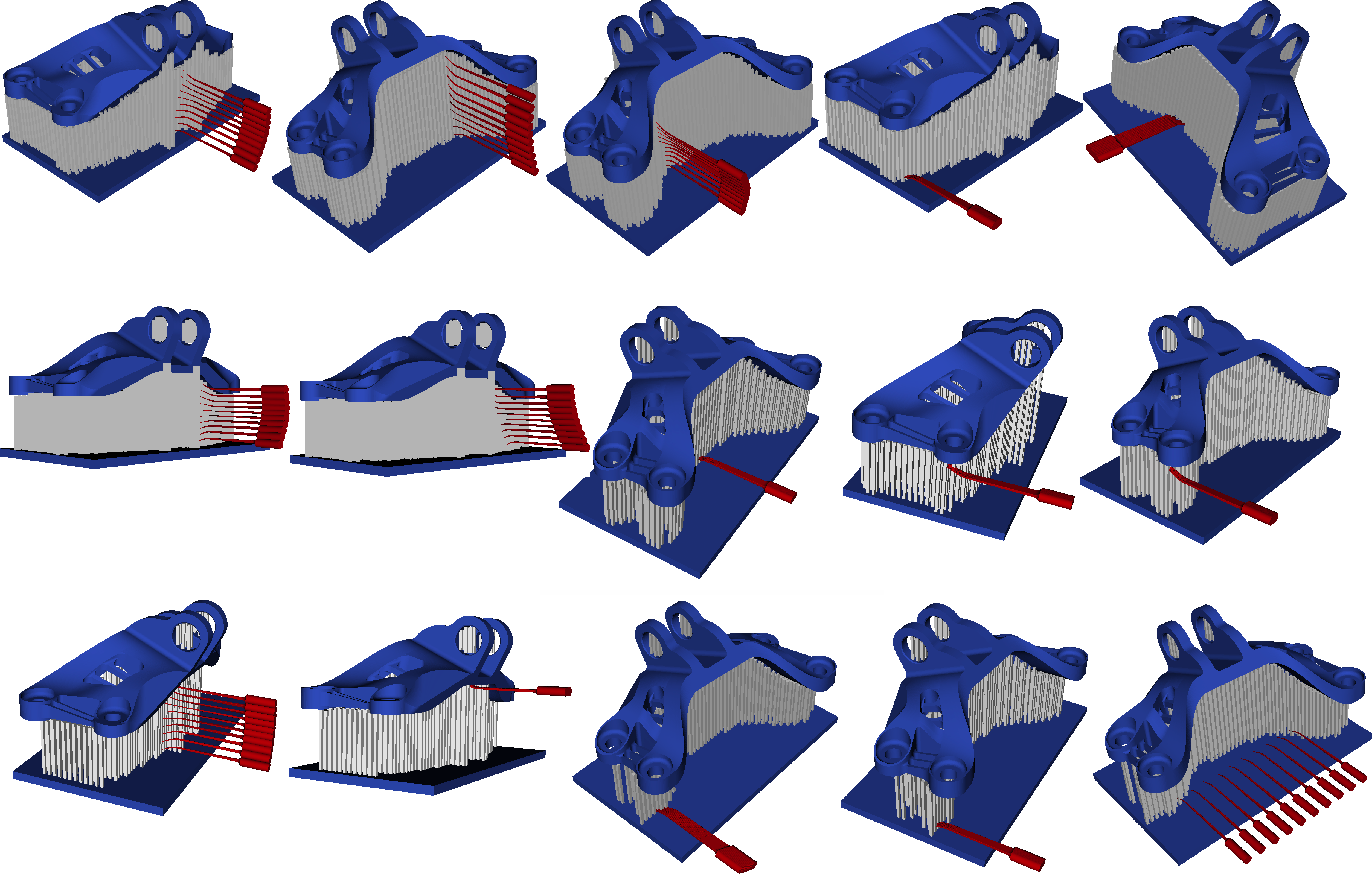}
	\caption{Several intermediate collision-free paths for the cutting tool to
		contact the supports at dislocation features. Each row represents a single
		recursion of Algorithm \ref{alg_recursive}. Observe that the motion planner
		ensures that the tool changes orientation if required when moving from one
		dislocation feature to the next. Each individual collision free path is
		computed using \textsf{OMPL} \cite{Sucan2012open} in approximately 0.5 seconds}.
	\label{fig_plan}
\end{figure*}

We demonstrate the approach using an illustrative 3D example. A near-net shape
along with a cutting tool to remove supports at dislocation features is shown in
Fig. \ref{fig_setup}.

We assume that the part is fixtured properly (not shown here, similarly to Fig.
\ref{fig_metal}) such that the removal sequence of the support components does
not impact its stability.

At every round, Algorithm \ref{alg_recursive} identifies the outer layer of
supports that can be peeled off, using FFT-based convolution (Sections
\ref{sec_obs} and \ref{sec_contact}) as illustrated in Fig. \ref{fig_forest}.
For each layer, solving the TSP on the graph of fibers (Section \ref{sec_plan})
gives a sequence of configurations to visit, while asserting all of them will
contact a distinct dislocation feature without collision.

The \textsf{OMPL} motion planner is invoked to find a collision-free tool-path
to fracture the dislocation features one after another (Section \ref{sec_ompl}),
as illustrated in Fig. \ref{fig_plan}.

Figures \ref{fig_obs} and \ref{fig_tsp} show running times for computing
$\conf-$space obstacles and solving TSP, respectively.
\section{Conclusion}

Support removal is an important post-processing step for AM and is often a
bottleneck for  manufacturability. We have shown an automated solution to
identifying and removing support materials in an AM part, by exploiting the
properties of the $\conf-$space obstacle, free space, and contact space obtained
as their shared boundary. The contact space is extracted as an interval level
set of a sparse field that is precomputed and queried during a recursive
algorithm to peel off removable supports. The approach does not make
assumptions on part, tool, or support structure geometries, or on the DOF with
which a tool can move. \com{We demonstrate several non-trivial three dimensional
	examples to illustrate our approach.}

Several improvements to this paper are possible. The computations are highly
parallelizable and offer opportunities for more efficient implementation.
Furthermore, uniform sampling of the rotation space for $\conf-$space
computations can be improved by choosing rotations based on more specific
manufacturing constraints.

As opposed to explicitly computing $\conf-$space obstacles and finding goal
states for a motion planner, one may directly find collision-free paths between
neighboring dislocation features by locally sampling configurations and invoking
rapid collision-checkers. This is similar to the probabilistic roadmap approach
to path planning
\cite{Kavraki1994probabilistic,Kavraki1996probabilistic,Hsu98finding,Boor1999gaussian}.
Nonetheless, this paper demonstrates that explicit calculation of 6D
$\conf-$space maps should not be considered as an impediment to solving
practical mechanical design and manufacturing problems. We hope that it inspires
researchers to apply these methods to other practical problems in which the
specific properties of the application can be leveraged to make $\conf-$space
computations tractable.

Last but not least, while this paper's focus has been kinematics and spatial
planning aspects, there are many unsolved problems with regard to the physics
of support removal (e.g., mechanics of the fracture process) that have been
ignored and need to be considered in practice.

\section*{Acknowledgement}

This research was developed with funding from the Defense Advanced Research
Projects Agency (DARPA). The views, opinions and/or findings expressed are those
of the authors and should not be interpreted as representing the official views
or policies of the Department of Defense or U.S. Government.


\bibliographystyle{elsarticle-num} 
\bibliography{suppRem}

\end{document}